# An analysis of visitors' length of stay through noninvasive Bluetooth monitoring in the Louvre Museum

Yuji Yoshimura[a], Anne Krebs[b], Carlo Ratti[a]

[a] SENSEable City Laboratory, Massachusetts Institute of Technology, 77 Massachusetts Avenue, Cambridge, MA 02139, USA;

[b] Dominique-Vivant Denon Research Centre, muse du Louvre, 75058, Paris, Cedex 01, France;

**Abstract:** Art Museums traditionally employ observations and surveys to enhance their knowledge of visitors' behavior and experience. However, these approaches often produce spatially and temporally limited empirical evidence and measurements. Only recently has the ubiquity of digital technologies revolutionized the ability to collect data on human behavior. Consequently, the greater availability of large-scale datasets based on quantifying visitors' behavior provides new opportunities to apply computational and comparative analytical techniques. In this paper, we attempt to analyze visitors' behavior in the Louvre Museum from anonymized longitudinal datasets collected from noninvasive Bluetooth sensors. We examine visitors' length of stay in the museum and consider this relationship with occupation density around artwork. This data analysis increases the knowledge and understanding of museum professionals related to the experience of visitors.

## Introduction

The recent development of emerging technologies and their rapid diffusion into our daily life has caused a structural change in human behavior analysis. Indeed, the ubiquitous presence of wired and wireless sensors in contemporary urban environments produces an empirical record of individual activities at detailed levels. In addition, to the ubiquity of this technology, computationally advanced computer systems make accumulating large datasets of human behavior at high frequencies possible-sometimes even in real time.

Contrary to the common use of such data-collection technology, the data collection of visitors' behaviors in large-scale art museums has not advanced much over the past few decades. The traditional pencil-and-paper based tracking method is still widely used in the form of "timing and tracking" [1]. Furthermore, many of the emerging technologies don't work appropriately in the museum setting for several reasons: active mobile phone tracking with GPS [2] doesn't work inside buildings and passive mobile phone tracking [3,4] cannot distinguish visitors' presence and movement between rooms because its detection range, based on the antenna's coverage, is too large. Video camera based tracking technologies are useful, but the substantially higher cost do not allow for the necessary infrastructure in a museum environment. RFID [5], ultra wideband [6] and mobile phone centered wifi tracking method are promising, but they require visitors to be equipped with certain devices or to download the proper application in advance. This participatory process prevents us from generating large-scale datasets, which are necessary for relevant analysis.

For this purpose, this paper employs a Bluetooth detection technique [7-10]. Bluetooth detection is unobtrusive, making use of the visitors' digital footprint they unconsciously leave behind [11]. Furthermore, these datasets can be anonymized in order to protect the users' personal data. As a result, this technique enables us to generate large-scale datasets. Additionally, we can expect datasets to be obtained without any behavioral bias. Participants might adapt their behaviors if they become conscious of being observed. Furthermore, we can perform data collection for a longer period than just one day or a few days, which is a typical data collection technique, to form a hypothetical visitor, who can represent a whole population.

Thus, our methodology is to analyze "real" and large-scale empirical data, which is contrary to that of traditionally employed manual-based methods. This proposed method sheds light on unknown aspects of visitors' behaviors in terms of their length of stay and the influence of the crowd over a visitor.

**Bluetooth tracking system in the Louvre Museum**

Bluetooth detection systems are widely used to track people inside buildings [10], in urban settings [9], and to generate traffic information based on detecting vehicles [12]. Bluetooth detection systems work as follows: once a Bluetooth-activated mobile device enters the detectable area, the sensor continues to receive the emitted signal from the mobile device until the signal disappears. Thus, the sensor registers the time at which the signal of a mobile device appears, also called check-in time. Afterward, when the signal of a mobile device disappears, the sensor records the check-out time. Then the difference between each mobile device's check-in and check-out time can be calculated, which defines the length of stay at the node. Similarly, by looking at the first check-in time and the last check-out time over all nodes, it is possible to calculate how long a visitor stays in the Denon wing of the Louvre museum. Such a series of check-in and check-out time data registered by all the installed sensors makes it possible to construct a visitor's trajectory through the Denon wing of the museum, including their travel time between nodes. All of this information can be collected without invading visitor' privacy because Secure Hash Algorithm (SHA) encryption [13] is applied to each sensors by converting each device's MACID in to a unique identifier [7].

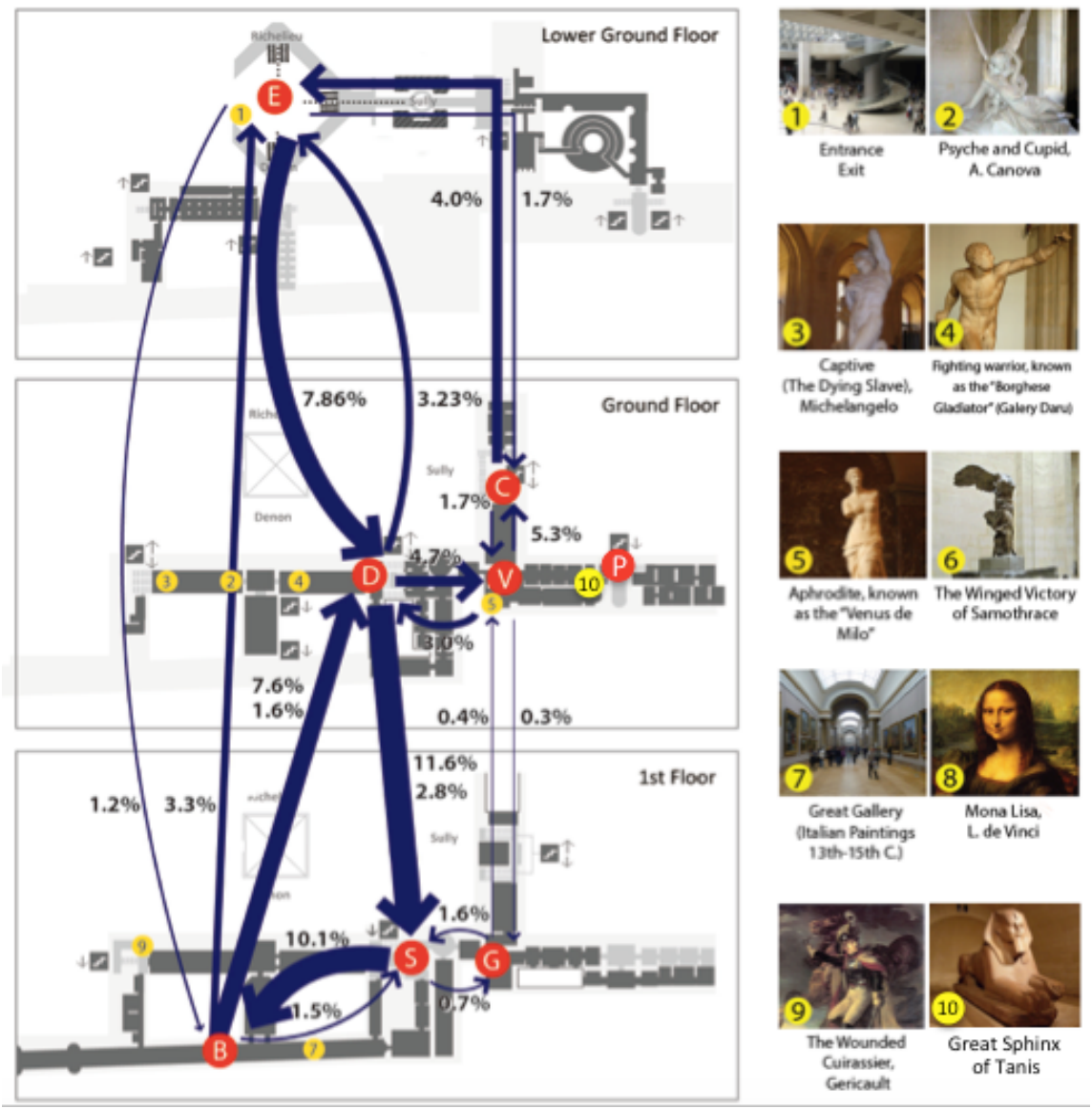

**Figure 1.** Location of eight sensors (E, D, V, B, S, G, C, P) indicating their approximating sensing range. The arrow and its width represents visitors' flow between nodes.

Eight Bluetooth sensors have been deployed throughout the Denon wing of the museum, covering key places to capture visitors' behavior. These sensors were placed along one of the busiest trails identified by Louvre Museum, which lead visitors from the entrance to the Venus de Milo. The sensors are placed at the Entrance Hall (E), Gallery Daru (D), Venus de Milo (V), Salle des Caryatides (C), Sphinx (P), Great Gallery (B), Victory of Samothrace (S), and Salle des Verres (G). The data collection was performed at different periods by a different number of sensors during a five-month period from April 2010 to August 2010. After data cleanup and data processing, which adjusted the data to remove any inconsistencies, 81,498 unique devices were selected to be analyzed for this paper. By comparing the number of detected mobile devices and ticket sales, we found that, on average, 8.2% of visitors had activated Bluetooth on their mobile phone. Additionally, we previously uncovered visitors' transition probability between nodes and their mobility patterns considering their length of stay in the museum (see Figure 1). Based on this previous study, this paper analyzes visitors' length of stay in each node as well as the entire duration of their visit, which is how this study differs from the previous one.

**Analyzing three factors regarding length-of-stay in the museum**

We analyze three different factors related to visitors' length of stay in the museum. The first factor relates to entry time, which can be used to assess the distribution of visitors' length of stay in the museum depending on when they entered the museum. The second factor provides visitors' length of stay at each specific node. The third factor is the relationship between the length of stay at a specific node and the number of visitors around the node (i.e., density).

*The length of stay in the museum*

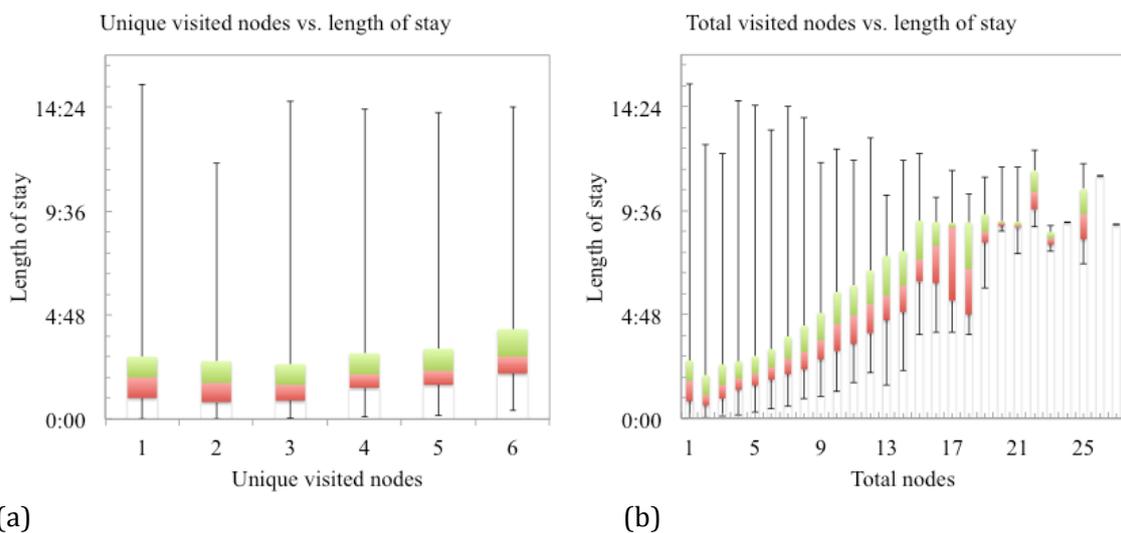

(a)                                            (b)
**Figure 2.** (a) The number of unique node visited against the length of stay (b) The number of total visited nodes against the length of stay.

Figure 2 (a) shows that the median length of stay is very similar across all amounts of unique visited nodes. We used a non-parametric correlation analysis (Spearman's rank correlation coefficient) because the variables do not seem to follow a normal distribution. We also include a series of boxplots to better explain the relationship between variables. A very low correlation value ($\rho = 0.073$, $p<2.2e-16$) suggests that there is no relationship between these two variables. That is, the number of unique nodes visited seems to be independent of the length of stay. On the other hand, Figure 2 (b) shows a different relationship between the length of stay and the total amount of visited nodes. The correlation coefficient ($\rho = 0.186$, p-value$<2.2e-16$) suggests a weak association between the two variables. Our interpretation is that when people stay for a longer period inside the museum, they tend to limit the number of visited nodes and prefer to dedicate more time to exploring those they visit thoroughly (sometimes visiting them more than once), instead of visiting a higher number of different nodes.

We also examine a distribution of an average of visitors' length of stay in the museum classified by the hour of the day they visit the museum. This analysis reveals whether or not visitors' entry time affects their length of stay at the museum. We divide a day of the week into two groups depending on the closing time of the museum. The first group consists of Monday, Thursday, Saturday and Sunday, when the door closes at 18:00 and the second group includes Wednesday and Friday, when the museum closes at 21:45.

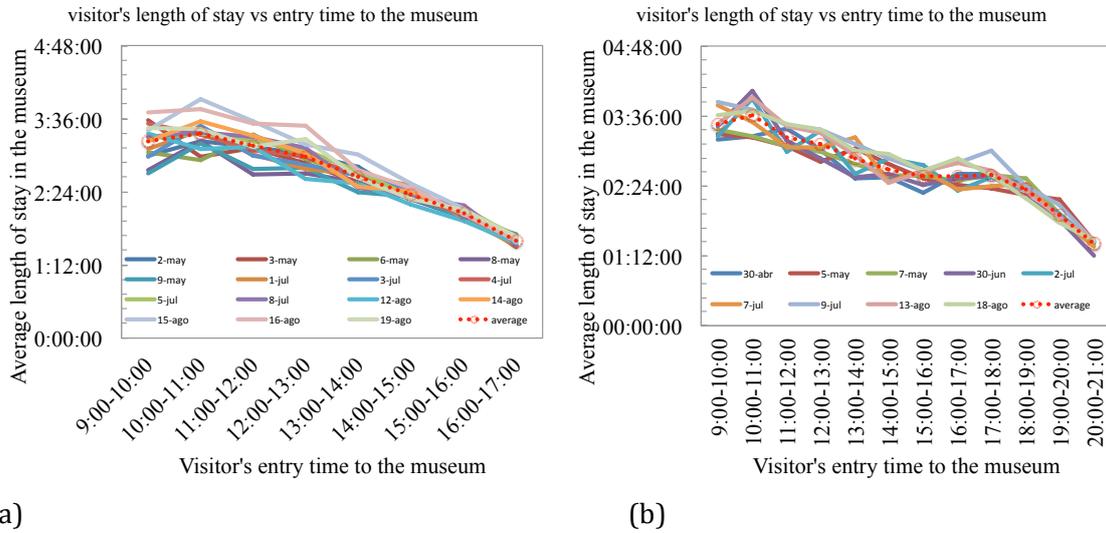

(a) (b)

**Figure 3.** (a) The distribution of the average stay time by visiting hour on Monday, Thursday, Saturday and Sunday. (b) The distribution of the average stay time by visiting hour on Wednesday and Friday.

Figure 3 (a) presents a clear tendency that the length of stay time decreases toward the closing hours of the museum. The earlier a visitor enters the museum, the longer that visitor tends to stay in the museum. This indicates that the closing time of the museum works as a constraint to limit the length of visitors' time in the museum. Thus, visitors seem to stay longer within their limited available time in order to maximize their benefits.

Conversely, the results on Wednesday and Friday show a different tendency of visitors (see Figure 3 (b)). As with the previous analysis, we can observe a tendency that the length of stay at the museum decreases with the advance of the time. However, the decrease in length of stay is slightly mitigated in the middle of the day. Just after the opening of the museum (i.e., 10:00-11:00), the length of stay is greatest, but in the late afternoon (i.e., 17:00-18:00), the length of stay increases slightly. This data makes us suppose there might be two kinds of visitors. That is, while some intend to maximize their utility (e.g., staying time) within the limited time the museum is open by visiting earlier, others try to take advantage of the longer hours and wait until the evening to visit.

All of these analyses and results indicate that the time visitors enter can be used to predict visitors' length of stay in the museum, but their length of stay in the museum doesn't have any correlation with the number of visited nodes over the course of their visit. While the longer length of stay slightly suggests a larger number of visited nodes, their relationship is not significant. Such a result is particularly useful for the daily management of peak periods and rush hours by the staff.

*The length of stay at each node*

We examine how long a visitor stays at each node in a disaggregated way and uncover the feature of each node by comparing the analysis of an accumulation of individual visitor's behavior.

Our preliminary analysis shows that node E and node S have a much longer length of stay than the other nodes. The median length of stay at nodes E and S are 00:16:29 and 00:19:03, respectively, while the average length of stay at the other nodes is 00:03:14. Node E is situated in the ticket sale desk, indicating, unsurprisingly, that visitors may wait for a long queue in order to purchase a ticket. Winged Victory of Samothrace, where sensor S is located, is one of the most attractive exhibits in the museum-together with the huge staircase in front of the exhibit. Besides, many visitors use the stairs as a improvised chair to take a rest during their visit. Those two factors make node E and node S different from other nodes. We speculate that the unique uses for these spaces results in their much longer length of stay, and this fact motivates us to exclude these nodes from our following analysis.

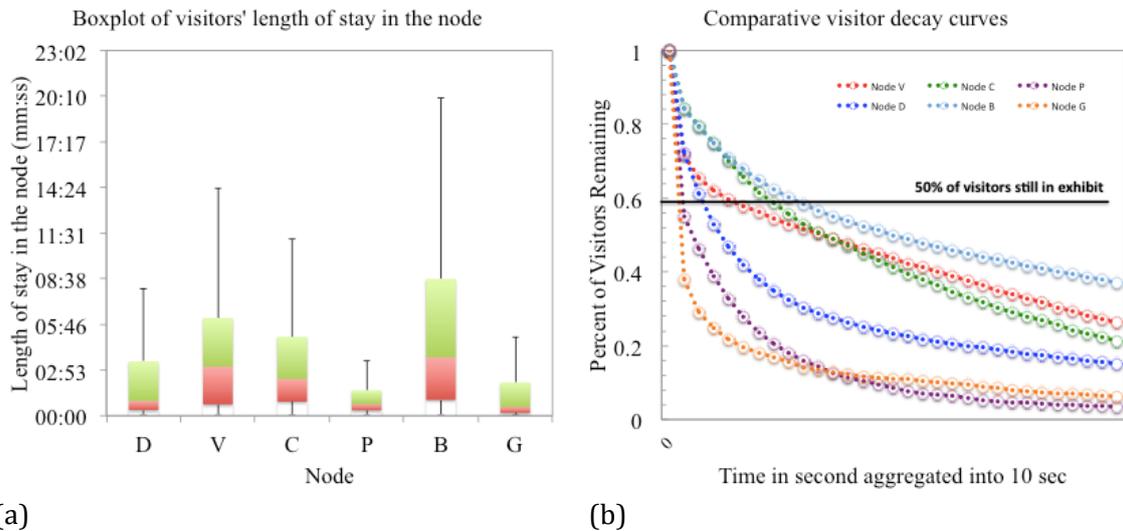

(a) (b)
**Figure 4.** (a) The boxplot of the length of stay in each node. (b) The comparative visitor decay curves.

Figure 4 (a) presents the boxplot of visitors' length of stay at each node. We can observe that two groups exist: node V (Venus de Milo), node C (Salle des Caryatides), and node B (Great Gallery) experience a longer stay; node D (Gallery Daru), node P (Sphinx), and node G (Salle des Verres) experience a shorter stay. While the median length of stay for the former group is 03:02, the median for the latter is 00:44. Among these nodes the visitors' length of stay and its range in node P is much shorter than that of other nodes, and as a result, 98% of all visitors to node P have a length of stay between 19 seconds and 113 second.

Conversely, Figure 4 (b) shows the comparative "visitor survival curve". We plot the percentage of visitors at given times. This plot is frequently used in visitor studies in order to analyze the length of stay, when half of the visitors leave, an exhibit or room (Bicknell, 1995). We can observe that the length of stay is largely varied among nodes: 00:20-00:30 for node D, 01:30-01:40 for node V, 01:30-01:40 for node C, 00:00- 00:10 for node P, 02:10-02:20 for node B, and 00:00-00:10 for node G. However, most of the nodes experience a length of stay within a few minutes. Again, we can observe that two kinds of nodes exist, which we can classify as shorter and longer stay type nodes. Within the former group, half of visitors left the node by 10-20 seconds in the cases of nodes G and P. In the case of the node D, this duration is little bit longer, but half of visitors stayed for just 20-30 seconds.

*Relationship between visitors' length of stay and density*

The perspective on length of stay, however, greatly changes when examining visitors' duration of stay in relationship to the degree of the occupancy of each node. Figure 5 shows the relationship between each node's occupancy normalized by the maximum number of visitors in the area (x-axis) and the average duration of stay expressed by seconds (y-axis). As we can see, a clear tendency exists among all data. The average duration of stay first goes up with the room occupancy from point W to point X, then stays around the maximum on some occupancy level interval (point X to Z). After that, length of stay drops down as the occupancy level starts to exceed a certain threshold (point Z).

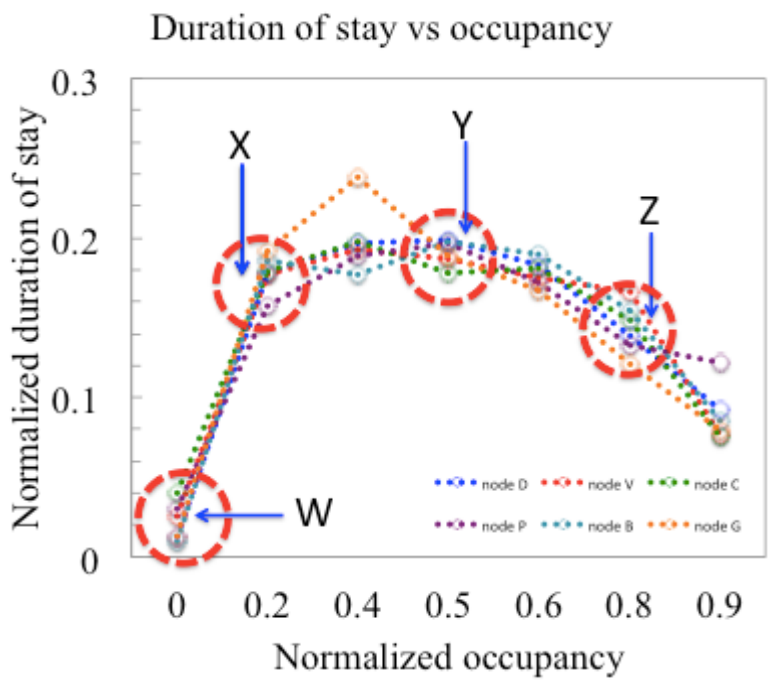

**Figure 5.** Distribution of the normalized occupancy vs the length of stay.

**Table 1.** The threshold of the normalized occupancy in each node.

|         | Point X        | Point Y        | Point Z        |
|---------|----------------|----------------|----------------|
| *Node D* |                |                |                |
|         | 183sec (0.235) | 204sec (0.498) | 188sec (0.636) |
| *Node V* |                |                |                |
|         | 289sec (0.235) | 315sec (0.368) | 271sec (0.781) |
| *Node C* |                |                |                |
|         | 225sec (0.221) | 249sec (0.352) | 187sec (0.753) |
| *Node P* |                |                |                |
| Node B | 94sec (0.214) | 118sec (0.5) | 79sec (0.764) |
|  | 353sec (0.238) | 375sec (0.504) | 293sec (0.772) |
| Node G | 167sec (0.267) | 208sec (0.351) | 145sec (0.657) |

The length of stay for point W is extremely short because it contains visitors, who just pass by the area rather than stay there, which we demonstrated in the previous sections. When the occupancy level increases from point W to X, the number of visitors who tend to stay longer also increases. Point X can be considered the equilibrium between an ideal length of stay for a visitor when he or she is free to stay as long as he or she wants. This is because visitors are free to look at the artwork between point W and X without any obstacles due to the low density of other visitors. Conversely, from point X onward, the average length of stay remains almost flat until point Z, at which point visitors' length of stay drastically starts to decrease. We speculate that this is because the exceeding high-density of other visitors may affect a visitor's comfort, resulting in a desire to escape the crowd.

The length of stay at point X is largely varied depending on the node. Node B at point X shows the longest length of stay (353 seconds) with a 0.238 normalized occupancy level among the other nodes. Then, node B remains a flat until point Z. Conversely, node P at point X represents the shortest length of stay (94 seconds) with a 0.214 normalized occupancy level among the other nodes. Although the occupancy level is similar between those two nodes, the former's length of stay is almost four times longer than the latter's. However, the normalized occupancy level of point X for node B is higher than that of node P (i.e., 0.238 vs. 0.214, respectively). Additionally, these nodes correspond with the maximum and minimum lengths of stay at point X, although the occupancy level of both nodes is quite similar: both of them are around 0.50. Furthermore, the length of stay at node P starts to decrease earlier than that of node B.

Regarding the relationship between node V and node B, although point X for both of them shows a similar length of stay (i.e., 0.235 vs. 0.238, respectively), the maximum length of stay of node V has a much lower occupation level than those of nodes P and B (i.e., 0.368 for node V). Additionally, node V has the highest density for point Z (0.781) with the second longest length of stay (271sec), whereas the longest length of stay is at node B (293seconds).

All of these facts indicate that the artworks at node B attract and hold visitors much more strongly than the artworks of node V and P. Node B seems to inspire visitors to stay longer, even with higher occupation density, while node V and P seem to cause visitors to stay for shorter durations when experiencing the same density. Also, the data shows that the average length of stay of a visitor and the occupation level of a node form a clear pattern. The crowd density around the artwork largely affects a visitor's length of stay either positively or negatively, and the type of effect is largely dependent on the node.

We speculate that, although it also depends on the nature of the galleries and the type of visitors, up to certain occupancy limits visitors are actually attracted by the crowd: however, once the crowd's size reaches a certain level, visitors will try to avoid the node. This indicates that we may use point Z as a threshold to distinguish visitors' level of comfort, which enables us to manage the environment in order to avoid exceeding this density threshold in the environment. Also, we may consider some characteristic points such as point X and Y to enabling to manage different types of environments/crowds inside the galleries.

**Conclusions and discussion**

The limitation of our proposed method is as follows: first, the sensor can detect only mobile devices in which the Bluetooth function is activated. This indicates the representativeness of the sample may have a strong bias toward certain groups, such as the upper class, higher-educated people, and the younger generation rather than seniors and children. Moreover, sensors are not capable of distinguishing individual visit or group visits. The representativeness of our sample can be calculated by comparing the sample obtained from Bluetooth detection and the head count by hand over a shorter period (e.g, a few hours or few days). In fact we applied a systematic comparison over a longer period (one month) with the number of ticket sales in the desk where a Bluetooth sensor is installed. Second, the sensor enables us to detect visitors' presence in the specific area, which is determined by the radius of each sensor's detection range, but such sensors cannot specify whether visitors are actually looking at the artwork or if they are simply in the area. This shortcoming is strongly related to the following limitation: "while time is a necessary condition for learning, time in a gallery does not correspond directly to time spent attending to exhibitions" [14]. Third, Bluetooth detection techniques cannot disclose the visitors' motivations and inner thoughts in any way: this method merely identifies their presence and the precise length of their stay. Finally, sensors are not capable to collect socio-demographics (i.e., origin, age, gender, profession) as for other traditional behavioral variables.

As a backdrop to this situation, the proposed new approach enables us to shed light on some unknown aspects of visitors' behavior. Our proposed system and the results we obtained through an adequate statistical analysis can work as a new tool for the museum management. Our system effectively captures an individual visitor's length of stay in different ways and enables us to store this information as a large-scale dataset. Such a method appears to be particularly useful for medium or small size museums where it could be possible to install a complete system of sensors. This method results in a dataset that is very different from the conventional dataset in two ways: the quantitative size of the dataset and the finer-grain detection of visitor behavior both spatially and temporally.

All of our findings are helpful for the management of visitor flow in order to reduce congestion at specific areas and around specific pieces of artwork. It is also useful in order to bring to light less visited or less "attractive" artworks/rooms inside the museum, in order to propose interpretation tools and/ or walking tours capable to increase the value of such "neglected" spaces. Additionally, our data suggest that visitor behavior is based on some patterns, which make it possible to foresee their future movement in a dynamic way. Also, these data are significantly useful for designing the spatial arrangement (e.g., changes in the layout of exhibits, facilities and advertisements), depending on visitor activities and use of space. Finally, our findings indicate that efficient and effective congestion management of the museum can be realized by limiting the number of visitors that are able to enter based on the time of the day.

**References**

1. Yalowitz, S. S. & Bronnenkant, Kerry. (2009) "Timing and Tracking: Unlocking Visitor Behavior" *Visitor Studies* 12(1): 47-64.
2. Shoval N, McKercher B, Birenboim A, Ng E, 2013, "The application of a sequence alignment method to the creation of typologies of tourist activity in time and space" Environment and Planning B: Planning and Design advance online publication, doi:10.1068/b38065
3. González M C, Hidalgo C A, Barabási A L, 2008, "Understanding individual human mobility patterns" *Nature* **453** 779-782
4. Ratti C, Pulselli R, Williams S, Frenchman D, 2006, "Mobile Landscapes: using location data from cell phones for urban analysis" *Environment and Planning B: Planning and Design* **33**(5) 727-748
5. Kanda T, Shiomi M, Perrin L, Nomura T, Ishiguro H, Hagita N, 2007, "Analysis of people trajectories with ubiquitous sensors in a science museum" *Proceedings 2007 IEEE International Conference on Robotics and Automation (ICRA'07)* 4846-4853
6. Tröndle, M, Greenwood, S, Kirchberg, V, Tschacher, W, 2014, "An Integrative and Comprehensive Methodology for Studying Aesthetic Experience in the Field: Merging Movement Tracking, Physiology, and Psychological Data", *Environment and Behavior*, 46 (1) pp102-135.
7. Yoshimura Y, Sobolevsky S, Ratti C, Girardin F, Carrascal J P, Blat J, Sinatra R, 2014, "An analysis of visitors' behaviour in The Louvre Museum: a study using Bluetooth data" *Environment and Planning B: Planning and Design* **41** (6) 1113-1131
8. Kostakos V, O'Neill E, Penn A, Roussos G, Papadongonas D, 2010, "Brief encounters: sensing, modelling and visualizing urban mobility and copresence networks" *ACM Transactions on Computer Human Interaction* **17**(1) 1-38
9. Versichele M, Neutens T, Delafontaine M, Van de Weghe N, 2011, "The use of Bluetooth for analysing spatiotemporal dynamics of human movement at mass events: a case study of the Ghent festivities" *Applied Geography* **32** 208-220
10. Delafontaine M, Versichele M, Neutens T, Van de Weghe N, 2012, "Analysing spatiotemporal sequences in Bluetooth tracking data" *Applied Geography* **34** 659-668
11. Mayer-Schönberger V, Cukier K, 2013, *Big Data: A Revolution That Will Transform How We Live, Work and Think* (John Murray, London)
12. Barceló J, Montero L, Marqués L, Carmona C, 2010, "Travel Time Forecasting and Dynamic Origin-Destination Estimation for Freeways Based on Bluetooth Traffic Monitoring" *Transportation Research Record: Journal of the Transportation Research Board* **2175**: 19-27
13. Stallings W, 2011, *Cryptography and Network Security: Principles and Practice, 5th Edition* (Prentice Hall, Boston MA)
14. Hein G, 1998, *Learning in the Museum* (Routledge, London)